\documentclass[aps,prl,twocolumn,groupedaddress,showpacs,english]{revtex4}
\usepackage[T1]{fontenc}
\usepackage[latin9]{inputenc}
\usepackage{amsmath}
\usepackage{amssymb}
\usepackage{graphicx}

\usepackage{caption}
\usepackage{wasysym}
\usepackage[version=4]{mhchem}
\usepackage{collcell}


\usepackage{babel}
\usepackage{chngcntr}
\begin{document}
\title{Coupling electrodynamic fields to vibrational modes in helical structures}

\author{Asaf Farhi and Aristide Dogariu}
\affiliation{CREOL, University of Central Florida, Orlando, Florida, USA 32816}
\begin{abstract}
Helical structures like alpha helices, DNA, and microtubules have profound importance in biology. It has been suggested that these periodic arrangements of constituent units could support collective excitations similarly to crystalline solids. Here, we examine the interaction between such constructs and oscillating dipoles, and evaluate the role of the helicity in the coupling between electrodynamic fields and vibrations. Based on a  vibrational and eigenfunction analyses we discover a group of modes of coherent oscillations that give rise to a strong and delocalized response, selectivity in  frequency, and typical interaction range. To describe the field scattering due to the structure vibrations we consider an anisotropic permittivity with a helical periodicity, which applies to all vibration types and close dipole locations. This new type of resonances identified here may help explain the role of electrodynamic fields in the diverse functionality of cytoskeletal microtubules in the cellular environment.

\end{abstract}

%
\maketitle

Microtubules (MTs) are tubular helical structures that self-assemble from their constituent tubulin-protein units. MTs are critical for the development and maintenance of the cell shape, transport of vesicles and other components throughout cells, cell signaling, and mitosis. Tubulins have a large dipole moment \cite{1,2,3,4} and it was conjectured that MT vibrations could generate electric field in its vicinity [5-7], also beyond the typical Coulomb and vdW range. This process can be powered by GTP hydrolysis, motor proteins that move along the MT, and mitochondria energy release [6]. MTs were also analyzed in the context of robust-edge topological vibrational modes [8], vibrational modes of hollow-cylinders, and two-dimensional (2D) crystal lattices [9-10]. Recently, their 3D mechanical vibrations were calculated numerically using a molecular structural-mechanics model [11] and their acoustic modes were measured experimentally under the assumption of thermal equilibrium [12]. Importantly, alternating electric fields were shown to inhibit cancerous cell-growth by an anti-MT mechanism \cite{kirson2007alternating}.

MTs have a highly regular helical shape that is rare in nature, similarly to carbon nanotubes \cite{13}. Their constituent units are identical, even more than in DNA and alpha helices, whose elementary units have different residues. In addition, the MT structure appears like a shifted crystal, which may give rise to axial propagation of vibrations. It is certainly of interest to understand how this exquisite geometry may affect oscillatory phenomena of MTs such as vibrations and electromagnetic (EM) excitations.  In a broader sense, one can ask if these properties are critical for the diverse functionality of MTs in biology. Of particular interest would be to understand the interaction with surrounding molecules and if the modes have a particular extent and frequency properties. 

In the following, we answer these questions to some extent. We first analyze vibrations in a helical structure by employing a top-view model that describes accurately macroscopic vibrations of complex structures. We then develop an eigenfunction analysis \cite{14,15,16,17} for the vibrational-mode-mediated interaction between a MT and an oscillating electric dipole in a host medium. To that end, we consider an infinitely-long dielectric structure consisting of units disposed in a helical arrangement. These units can vibrate in a collective manner and have internal vibrational and electronic excitations. A dipole in proximity to this structure emits radiation with a wavelength $\lambda=2\pi c/\omega\gg l,$ where $l$ is the typical length scale, and therefore the interaction can be analyzed in the quasistatic approximation \cite{32}. In this regime, the electric and magnetic fields are decoupled and the electric field, which oscillates in time, obeys Poisson's equation \cite{14,15,16}. 
We derive eigenfunctions that express the scattering of the electric field due to the vibrations. 
To descibe this interaction, we define the MT as an inclusion with a permittivity $\overleftrightarrow{\epsilon_1}(\omega,\mathbf{r})$ that is anisotropic and periodic along a helical orbit, and the host-medium with a permittivity $\epsilon_2(\omega),$ assuming $\omega>\mathrm{250MHz},$ in which ionic screening is negligible \cite{2}.


We consider a dipole that emits radiation, which impinges on the helical structure. In the near field, the dominant spatial frequencies of the emitted field correspond to wavelengths on the order of the distance from the dipole \cite{Pendry,16}. 
We define the incoming field $\mathbf{E}^\mathrm{inc}$ as the dipole field in a uniform $\epsilon_2$ medium \cite{14,15,16}. While this field is usually described with respect to the dipole position, we utilize its expansion with respect to the structure axis to relate it to the structure vibrations. 
We consider the case in which laterally-adjacent units move together, that for an axially-shifted crystal results in that each axial chain behaves as a 1D crystal.
To impose this movement, we require $\mathbf{E}^\mathrm{inc}$ to be symmetric to a continuous translation along a helical orbit. This situation is illustrated in Fig. 1 (a) for the case of a MT, in which the tubulin dimers are disposed in a helical arrangement. The electric fields in longitudinal and helical configurations are shown in Fig. 1 (b) and (c). As a result, the tubulin units can change their size and move as suggested in (d) and (e), respectively. In a helical-field arrangement, laterally-adjacent units move together (d). The movement where the adjacent tubulins are not aligned as shown in Fig. 1 (e) is assumed to be less favorable energetically.
\begin{figure}
\includegraphics[scale=0.7]{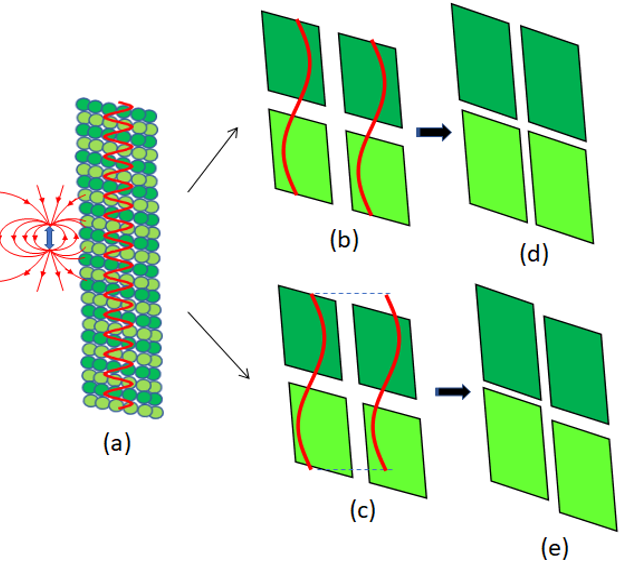}
\caption{The physical system: an oscillating dipole emitting electromagnetic field is in proximity to a MT. Part of the electric field couples to vibrational modes in the MT (a). The coupled field can be in a helical arrangement (b) or standard longitudinal arrangement (c). In response to the field, the tubulins deform and translate such that they move synchronously (d) or asynchronously (e).}
\end{figure}
Imposing this symmetry on $\mathbf{E}^\mathrm{inc}$ inside the structure results in (see Supplementary Material (SM) 1)
\begin{equation}
\mathbf{E}^\mathrm{inc}(\phi,z,\rho)\propto \exp\left(im\left(\phi-k_{z}z\right)\right)
\end{equation}
where $m$ is an integer number, $k_z=2\pi/a,$ $a$ is the helical-orbit axial periodicity, and $\phi,z,\rho$ are cylindrical-coordinates variables.
In these field modes, the $k$ and $m$  degrees of freedom are related by $k=mk_z$, which, if the medium responds strongly to them, implies selectivity in $\mathbf{k}$ and $\omega.$
Clearly, the high-$m$ modes have high spatial frequencies and can be dominant only for close dipole locations.
In addition, invariance of $\mathbf{E}^\mathrm{inc}$ to discrete lateral translations along the helix results in the same field distribution in each constituent unit and a coordinated movement. Such modes have high spatial frequencies $k_z n$, where $n$ is the number of units per helical round. These modes can be excited when the dipole is very close to the helical structure (typical interaction distance is $2\pi/k_z n$) and the field impinging on the structure has very high spatial frequencies.
 This situation is similar to the simpler case of the electrostatic field generated by charges in a helical arrangement with a uniform-inclusion permittivity \cite{18}.
 


Having analyzed the coupling of incoming EM fields to synchronous vibrations, we examine now in more detail the vibrations in the helical structure.
We first consider the forms of vibration. Radial movements are expected to be damped \cite{6} since they involve movements of a relatively large volume of liquid. While vibrations of a helical structure are different from vibrations of a spring, a spiral motion may also be less favorable mechanically since it involves movements of long helical chains. Moreover, in the context of MTs, the azimuthal dipole moment is small \cite{6} and in a recent work torsion was found to be insignificant \cite{11}. We will therefore focus on axial vibrations.

We now analyze classically the vibrational modes that can be excited by the incoming field and generate field. For 1D crystals, such a treatment agrees with the quantum analysis \cite{21}. We consider the coupling of vibrations also to field components with $kc\gg\omega$ that are almost static \cite{22}. While interaction of near field with a crystal was analyzed for $ka\ll1$ \cite{23,24}, we extend it to $ka\geq1.$ 
When vibrational modes and electric field are coupled they have the same $\omega,\mathbf{k},$ and at low and high $k\mathrm{s},$ $\omega(k)$ of one of the polaritons and the uncoupled vibrational mode are similar \cite{21}. We also show that in our case the 1D crystal symmetry $k\rightarrow k+k_z$ \cite{21} is not satisfied.
We consider a structure comprising two types of units with masses $m_1,m_2$ connected by springs $k_1,k_2,k_3,k_4$ as shown in Fig. 2 (b). Denoting the axial displacements of $m_{1,2}$ and the indices of the axial and lateral shifts by $u_{1,2}$ and $s,q$, respectively, and assuming $u_{1,2}=a_{1,2} e^{ikz+im\phi},$ we write the equations of motion (EOM)
\begin{widetext}
\begin{align}
-\omega^{2}m_{1}u_{1sq}=&k_{1}\left(u_{2sq}-u_{1sq}\right)-k_{2}\left(u_{1sq}-u_{2s-1q}\right)-k_{3}\left(u_{1sq}-u_{1sq+1}\right)-k_{3}\left(u_{1sq}-u_{1sq-1}\right),\nonumber\\
-\omega^{2}m_{2}u_{2sq}=&k_{2}\left(u_{1s+1q}-u_{2sq}\right)-k_{1}\left(u_{2sq}-u_{1sq}\right)-k_{4}\left(u_{2sq}-u_{2sq+1}\right)-k_{4}\left(u_{2sq}-u_{2sq-1}\right),
\end{align}
\begin{equation}
\left(\begin{array}{cc}
-\omega^{2}m_{1}+k_{1}+k_{2}+4k_{3}\sin^{2}\left(\left(ka/n-2\pi m/n\right)/2\right) & -\left(k_{2}e^{-ika}+k_{1}\right)\\
-\left(k_{2}e^{ika}+k_{1}\right) & -\omega^{2}m_{1}+k_{1}+k_{2}+4k_{4}\sin^{2}\left(\left(ka/n-2\pi m/n\right)/2\right)
\end{array}\right)\left(\begin{array}{c}
u_{1}\\
u_{2}
\end{array}\right)=\left(\begin{array}{c}
0\\
0
\end{array}\right).
\end{equation}
\end{widetext}
This \emph{1D description} enables us to analyze the behavior of the system in the axial axis while accounting implicitly for the lateral interactions in the terms with $k_3,k_4.$ These diagonal terms with $k_3,k_4$ restrain the movements of $m_1,m_2$ to their sites as in a local oscillator and vanish for the helical functions satisfying $k=mk_z$ (see Eq. (1)). These interactions are associated with propagation of axial movements along a helical orbit similarly to an infinite chain of identical particles \cite{21}.


For the $k=mk_z$ modes since $u_{1s,q+1}=u_{1s,q}e^{i\left(ka/n-2\pi m/n\right)},$ laterally adjacent units oscillate in-phase and form a standing wave, resulting in super-radiance and strong scattering in some cases \cite{25,26}, \cite{21} p. 102. In more complex structures the identical atoms and therefore the centers of mass of all the units move together, which relates this type of model also to low $\omega$/large mass vibrations. Eq. (3) can be written as $A\mathbf{u}=\omega^2 \mathbf{u},$ where A is a Hermitian matrix and therefore diagonalizable. Since $\omega^2$ is real and positive, we obtain $\omega(k),$ which means that the modes are delocalized. We now consider the response at a given $\mathbf{k}$ and therefore analyze the EOM at this $\mathbf{k}$. When anharmonicity or dissipation are incorporated, the matrix formulation and Hermiticity do not hold and localization can arise. We introduce anharmonicity in the axial forces between lateral units due to the alignment shift of the units upon movement. The sum of these (second order) forces $\propto u_{1sq}^{2}\left(1-\cos\left(ka/n-2\pi m/n\right)\right)$ and translates to an \emph{on-site} term. For the $k=mk_z$ modes these forces vanish and the $u_1-u_2$ coupling terms are maximal. Moving away from $k=mk_z$ increases the ratio of anharmonicity to dispersion, leading to a more localized response, similarly to interacting diatomic molecules with \emph{internal} anharmonicity \cite{27,28} (see SM 2.1). In the SM 2.2 we perform a similar analysis for two helical structures without axial periodicity and axial interactions and obtain similar properties. Such properties were recently observed in DNA [29]. We also analyze the effect of dissipation by introducing $\gamma \dot{u}_{1,2}$ terms, which shows that $\mathrm{Re}(\omega(k))$ is hardly affected and $\mathrm{Im}(\omega(k))$ is \emph{constant} at all $k\mathrm{s},$ except at large $\gamma\mathrm{s}$ that suppress the acoustic modes (SM 2.3). Since we consider axial vibrations and in Ref. \cite{9} the vibrations of a cylindrical-shell-water system behave similarly to a free shell for our $m\geq1$ modes, we assume that anharmonicity is a more dominant effect at least for the optical modes.

Interestingly, the $\alpha,\beta$ units of the MT have electrical charges with the same sign \cite{1}. This may imply that ``acoustic'' modes, for which adjacent units move together \cite{21}, generate current and couple to electric field along with optical modes. From Eq. (3) we calculate $\omega(k)$ for the acoustic and optical modes. We find that the $k=mk_z$ modes have the same $\omega(k)$ of a 1D crystal (see Fig. 2 (c) and SM videos) in agreement with the previous analysis in Eq. (1). In addition, one can substitute $\omega_T\rightarrow\omega(\mathbf{k})$ in the expression for $\epsilon$ \cite{21} and obtain 
$\epsilon(\omega,\mathbf{k})=1+4\pi Nq^2/[m_r(\omega^2(\mathbf{k})-\omega^2)],$ where $q$ is the charge, $m_r$ is the reduced mass, and $N$ is the charge concentration.

Moreover, the $k=mk_z$ field modes have the same potential distribution in each dimer and for a fixed dimer length (corresponds to internal vibrations or acoustic modes) the dimers can be treated as non-interacting also in the axial axis that may result in a similar spectrum for a dimer and the structure, which agrees with Ref. \cite{30}. 
\begin{figure}
\includegraphics[scale=0.52]{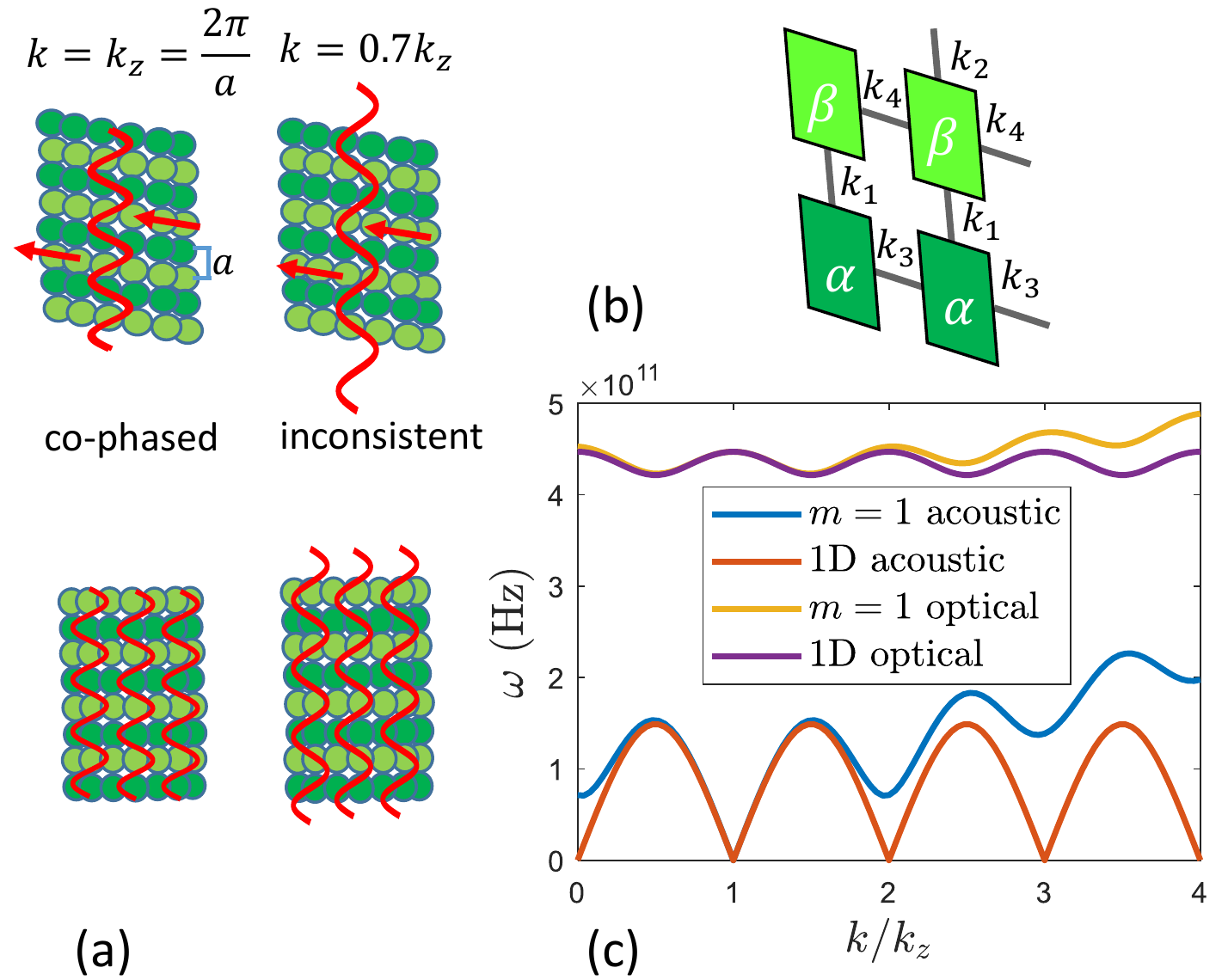}
\caption{Vibrational-mode analysis for a helical structure. The illustrations show that $k=k_z m$ are allowed when requiring decoupling between the axial protofilaments (a). The structure is composed of two units denoted by $\alpha,\beta$ with masses $m_1,m_2$ connected with springs $k_1,k_2,k_3,k_4$ (b). $\omega(k)$ for the acoustic and optical $m=1$ helix and 1D crystal modes. The MT parameters are $m_1=m_2=0.9\cdot10^{-22}\,(\mathrm{Kg}),k_1=8,k_2=1,k_3=k_4=2\, (\mathrm{N/m}),n=13,$ where $k_4$ is of the order of magnitude of the value in Ref. \cite{10}. (c) }
\end{figure}

Having described the vibrational modes of the helical structure, we now examine the scattering of the electric field due to these vibrations. To this end, we will use the eigenstates $\psi_k$ of the quasi-electrostatic potential. In a composite medium $\psi_k$ represents the potential of a field that exists without a source for an inclusion eigenpermittivity $\epsilon_{1k}.$ $\psi_k\mathrm{s}$ can be used to expand the scattered electric field $\psi_{\mathrm{sc}},$ which is generated due to the existence of the inclusion. In turn, $\epsilon_{1k}\mathrm{s}$ are calculated by imposing field boundary conditions and depend on the inclusion geometry. For propagating waves, this requires gain and constructive interference as in a laser. However, for evanescent waves $\epsilon_{1k}\mathrm{s}$ are real and can be reached more naturally. When the inclusion permittivity $\epsilon_1\approx \epsilon_{1k},$ a physical resonance occurs and the system responds resonantly \cite{14,15,16}. 

Let us describe $\psi_{\mathrm{sc}}$ for an anisotropic inclusion permittivity with a helical periodicity, which enables us to account for surface roughness. We first associate the permittivity \emph{tensor} to axial vibrations by considering an anisotropic inclusion with an axial permittivity $\epsilon_z (\mathbf{r})$ and $\epsilon_{\rho}=\epsilon_{\phi}=\epsilon_2.$ 
We then utilize the structure symmetries to analyze a permittivity with helical periodicity. In crystals, the permittivity is usually expanded in a Fourier series and it couples each field mode with the modes with $\mathbf{k}+\mathbf{G}_n,$ where $\mathbf{G}_n$ is a reciprocal-lattice vector, and there is an effective $\overleftrightarrow{\epsilon_1}(\omega,\mathbf{k})$ that describes the $\omega,\mathbf{k}$ response to an excitation at $\omega,\mathbf{k}$ \cite{14,22,32,33,34,35,36,37,38}.
In our case, the symmetry to discrete translations defines the $k=mk_z$ and $k=nk_z$ modes that represent the ``DC'' and higher-order Fourier components, respectively. Thus, the coupling is to modes with integer multiples of $(\Delta m,\Delta k)=(1,k_z)$ and $\Delta k=nk_z$ apart. This form of $\overleftrightarrow{\epsilon_1}(\omega,\mathbf{k})$ is justified for the MT because $\lambda/a\gg 1$ and $\rho_\mathrm{ext} (\omega)=0,   \mathbf{J}_\mathrm{ext}(\omega)=0$ inside the inclusion, since the charges oscillate only in response to external excitations \cite{22,39, 40}.

We now turn to the quantitative analysis of the dipole-helical structure interaction.
In the SM 3 we show that for $\epsilon_z (\mathbf{k})$ and $\epsilon_{\rho}=\epsilon_{\phi}=\epsilon_2,$ the amplitude of $\psi_k$ in the expansion of $\psi_{\mathrm{sc}},$ $C_{\mathbf{k}\omega}\propto\left(\epsilon_{2}(\omega)-\epsilon_{1z}\left(\boldsymbol{k},\omega\right)\right)/\left(\epsilon_{1z\boldsymbol{k}}-\epsilon_{1z}\left(\boldsymbol{k},\omega\right)\right)\int\theta_{1}d\boldsymbol{r}\partial\psi_{\boldsymbol{k}}^{*}/\partial z{E}_{\boldsymbol{k}z}^{\mathrm{inc}},$ where $\theta_1=1$ in the $\epsilon_1$ volume, and therefore $\boldsymbol{E}_{\boldsymbol{k}}^{\mathrm{inc}}$ results in a contribution of $\psi_{\mathbf{k}}$ with the same $\boldsymbol{k}$ in the expansion. Thus, we write the $\psi_m\mathrm{s}$ that describe the spatial dependency of $\psi_{\mathrm{sc}}$ due to the $k=mk_z$ vibrations 
\begin{equation}
\psi_{m}=e^{im\left(\phi-k_{z}z\right)}\left\{ \begin{array}{cc}
A_{1m}K_{m}(mk_z\rho) & \rho>\rho_{2}\\
A_{2m}I_{m}+A_{3m}K_{m} & \rho_{1}<\rho<\rho_{2}\\
A_{4m}I_{m}(mk_z\rho) & \rho<\rho_{1}  
\end{array}\right.,
\end{equation}
where $\rho_1,\rho_2$ are the internal and external inclusion radii, $K_m$ and $I_m$ are the modified Bessel functions. 
We then solve Laplace's equation in cylindrical coordinates in $\rho_1<\rho<\rho_2$ to find the argument of the functions 
\begin{equation}
\epsilon_{2}\frac{1}{\rho}\frac{\partial}{\partial\rho}\left(\rho\frac{\partial\psi_{m}}{\partial\rho}\right)-\epsilon_{2}m^{2}\frac{1}{\rho^{2}}\psi_{m}-k_{z}^{2}m^{2}\epsilon_{zm}\psi_{m}=0,
\end{equation}
and obtain $I_m(mk_z \sqrt{\epsilon_{1zm}/\epsilon_2}\rho)$ and $K_m(mk_z \sqrt{\epsilon_{1zm}/\epsilon_2}\rho).$
To simplify $C_{\mathbf{k}\omega}$ we show in the SM 3 that the integral in $C_{\mathbf{k}\omega}$  $\propto\epsilon_{2}(\omega)/\left(\epsilon_{2}(\omega)-\epsilon_{1z\boldsymbol{k}}\right)\ensuremath{\nabla\psi_{\boldsymbol{k}}^{*}\left(\boldsymbol{r}_{0}\right)\cdot\boldsymbol{p},}$ where $\mathbf{p}$ is the dipole moment, and $\mathbf{r}_0$ is the dipole location.

Let us now analyze the scaling of $\psi_m$ for small and large $\rho\mathrm{s}$. We first observe that the $m=0$ mode is constant everywhere and can therefore be omitted. This mode is, however, relevant in the far field. For an infinite cylinder, when $k=m=0$ and $\rho_2\ll \lambda,$ it has the form outside the structure for $k\rho\gg 1$ of
$E^\mathrm{TM}_{z  ,m=0}\propto\sqrt{k_0/\rho} e^{i(k_0\rho-\pi/4)} ,$  where $k_0=\omega/c$ \cite{19}. Interestingly, this mode extends far from the helical structure and scales as $\sqrt{k/\rho}$. Now we examine the scaling of the $m\geq1$ modes. For $\rho\gg a,$ $K_{m\geq1} (mk_z\rho)\rightarrow\frac{1}{\sqrt{2mk_z}} \sqrt{\frac{\pi}{\rho}} e^{-mk_z \rho},$ which means that the typical interaction distance for a dipole is of the order of $a$ from the structure. Inside the MT the modes scale as $\lim_{\rho\rightarrow0}I_{m}\left(mk_z\rho\right)\propto\left(\frac{mk_{z}\rho}{2}\right)^{m}.$ Importantly, these modes are discrete and are dominated by the $m=1$ mode for $\rho_0-\rho_2\gtrsim a/2,$ where $\rho_0$ is the dipole radius. This means that the response of these modes is highly selective in $\mathbf{k}.$ 
 For $\rho_0-\rho_2\gtrsim a/2$ the $m=1$ mode is excited by the dipole and couples to the high-order modes that have a negligible effect at $\rho=\rho_0$ and we can consider approximately only this mode. 
In Fig. 3 we present the radial dependence of the first modes outside a MT. The modes have a typical interaction distance of the size of $a=8\mathrm{nm},$ which is larger than the Debye distance of 1nm \cite{2}, and the $m=1$ 
mode dominates at large distances. These functions have $m$ in the radial argument unlike the standard cylindrical modes. Two isopotential surfaces of $\psi_{m=1} (\mathbf{r})$ outside the helical structure are shown in the inset.
\begin{figure}
\includegraphics[width=8cm]{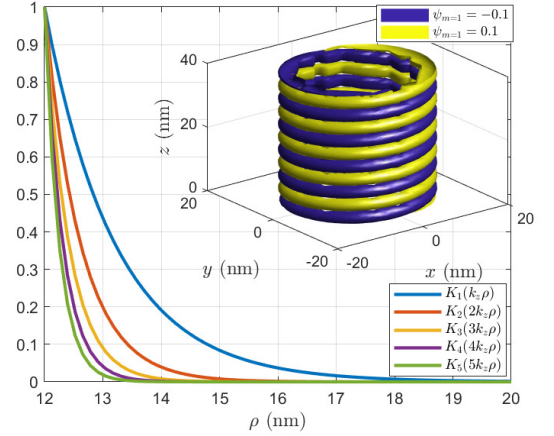}
\caption{Normalized $K_m (mk_z\rho)$  outside the microtubule. The interaction distance is of the order of $a.$ Inset: two isopotential surfaces of $\psi_{m=1} (\mathbf{r})=\pm 0.1$ extending to a radius of about 18nm.}
\end{figure}

To obtain the eigenpermittivities we impose the field boundary conditions (see SM 3). We also calculate $\mathbf{E}_{m\geq1}$ and find that $E_z>E_\rho>E_\phi,$ which means that the dipole tends to align almost parallel to the helical structure (see SM 4).


The resonances are approached when $\epsilon_{1zk} \approx\epsilon_{1z} (\mathbf{k},\omega_1),$ where $\omega_1$ is a resonance frequency.  Delocalization of modes implies $\mathrm{Im}(\omega_k)\approx 0$ and hence $\mathrm{Re}(\epsilon_{1z} (\mathbf{k},\omega))$ that spans over a large range of values. Therefore, close to $k=mk_z$ the system is likely to be at a resonance at $\omega_1\simeq \omega_{k=mk_z}.$ When exciting at $\omega_1$ there can be a strong and collective response \cite{41} of the helical structure that may affect the MT functionality. For resonances in additional types of structures see Refs. \cite{42,43,44,45}. $\epsilon_{1zk}$ depends on the structure dimensions and $\epsilon_{1zk}(\omega)$ on the internal interactions. Hence, $\omega_1$ may enable us to distinguish between different helical structures. 

The phenomena associated with the helical structure are both in the near and far fields (for $m\geq1$ and $m=0,$ respectively). They can be observed by absorption spectroscopy \cite{46} with an incoming field polarized along the axial axis \cite{47}, by Raman spectroscopy \cite{48}, or indirectly, by conductivity measurements \cite{30}.

In conclusion, we studied the coupling between EM fields and vibrational-modes in a helical crystal structure by analyzing the bulk and geometric properties of the structure. In particular, we examined the interaction of the structure and oscillating dipole, which emits field components also beyond the first Brillouin zone. We discovered a group of discrete modes of in-phase oscillations that give rise to a delocalized response and selectivity in $\omega$ and $\mathbf{k.}$ We note that in a recent experiment coherent and delocalized response was observed in DNA \cite{29}. We found that the first mode is long-range and scales as $1/\sqrt{\rho}$ while the other modes are quasistatic and have typical interaction distances characterized by the helical-orbit axial periodicity.
The fact that the spatial distribution of the $m=1$ mode correlates with the constituent units may imply spatial selectivity, which can be relevant for processes like self-assembly and induced polymerization. Finally, similar phenomena may arise in other physical systems where the constituent units are self-assembled \cite{49}.

 \nocite{*}
\bibliographystyle{plain}

\end{document}